\documentclass[a4paper,fleqn,usenatbib]{mnras}

\usepackage{newtxtext,newtxmath}

\usepackage[T1]{fontenc}
\usepackage{ae,aecompl}

\usepackage{graphicx}	
\usepackage{natbib,times}
\usepackage{array}

\title[The luminosities of Type-II SN progenitors]{The `Red Supergiant Problem': the upper luminosity boundary of  type-II supernova progenitors}

\author[B. Davies \& E. Beasor]{
Ben Davies,$^{1}$\thanks{b.davies@ljmu.ac.uk} and Emma R.\ Beasor$^{2,1}\thanks{Hubble Fellow}$
\\
$^{1}$Astrophysics Research Institute, Liverpool John Moores 
University, Liverpool Science Park ic2, \\ 146 Brownlow Hill, Liverpool, L3 5RF, UK\\
$^{2}$NSF's National Optical-Infrared Astronomy Research Laboratory, 950 N. Cherry Ave., Tucson, AZ 85721, USA
}

\date{Accepted 2020 January 16. Received 2020 January 16; in original form 2019 August 02 }

\pubyear{2019}

\begin{document}
\label{firstpage}
\pagerange{\pageref{firstpage}--\pageref{lastpage}}
\maketitle
      
\begin{abstract}
By comparing the properties of Red Supergiant (RSG) supernova progenitors to those of field RSGs, it has been claimed that there is an absence of progenitors with luminosities $L$ above $\log(L/L_\odot) > 5.2$. This is in tension with the empirical upper luminosity limit of RSGs at $\log(L/L_\odot) = 5.5$, a result known as the `Red Supergiant Problem'. This has been interpreted as evidence for an upper mass threshold for the formation of black-holes. In this paper, we compare the observed luminosities of RSG SN progenitors with the observed RSG $L$-distribution in the Magellanic Clouds. Our results indicate that the absence of bright SN II-P/L progenitors in the current sample can be explained at least in part by the steepness of the $L$-distribution and a small sample size, and that the statistical significance of the Red Supergiant Problem is between 1-2$\sigma$ . Secondly, we model the luminosity distribution of II-P/L progenitors as a simple power-law with an upper and lower cutoff, and find an upper luminosity limit of $\log(L_{\rm hi}/L_\odot) = 5.20^{+0.17}_{-0.11}$ (68\% confidence), though this increases to $\sim$5.3 if one fixes the power-law slope to be that expected from theoretical arguments. Again, the results point to the significance of the RSG Problem being within $\sim 2 \sigma$. Under the assumption that all progenitors are the result of single-star evolution, this corresponds to an upper mass limit for the parent distribution of $M_{\rm hi} = 19.2{\rm M_\odot}$,  $\pm1.3 {\rm M_\odot (systematic)}$,  $^{+4.5}_{-2.3} {\rm  M_\odot}$ (random) (68\% confidence limits).
\end{abstract}


\def\ga{\mathrel{\hbox{\rlap{\hbox{\lower4pt\hbox{$\sim$}}}\hbox{$>$}}}}
\def\la{\mathrel{\hbox{\rlap{\hbox{\lower4pt\hbox{$\sim$}}}\hbox{$<$}}}}
\def\msunyr{$M$ \mbox{$_{\normalsize\odot}$}\rm{yr}$^{-1}$}
\def\msun{$M$\mbox{$_{\normalsize\odot}$}}
\def\zsun{$Z$\mbox{$_{\normalsize\odot}$}}
\def\rsun{$R$\mbox{$_{\normalsize\odot}$}}
\def\minit{$M_{\rm init}$}
\def\mmax{$M_{\rm max}$}
\def\lsun{$L$\mbox{$_{\normalsize\odot}$}}
\def\mdot{$\dot{M}$}
\def\mdotdj{$\dot{M}_{\rm dJ}$}
\def\lbol{$L$\mbox{$_{\rm bol}$}}
\def\logl{$\log(L/L_\odot)$}
\def\kms{\,km~s$^{-1}$}
\def\EW{$W_{\lambda}$}
\def\arcsec{$^{\prime \prime}$}
\def\arcmin{$^{\prime}$}
\def\teff{$T_{\rm eff}$}
\def\Teff{$T_{\rm eff}$}
\def\logg{$\log g$}
\def\logz{$\log Z$}
\def\logl{$\log (L/L_\odot)$}
\def\vdisp{$v_{\rm disp}$}
\def\bcv{{\it BC$_V$}}
\def\bci{{\it BC$_I$}}
\def\bck{{\it BC$_K$}}
\def\lmax{$L_{\rm max}$}
\def\um{$\mu$m}
\def\chisq{$\chi^{2}$}
\def\AV{$A_{V}$}
\def\hminus{H$^{-}$}
\def\Hminus{H$^{-}$}
\def\ebmv{$E(B-V)$}
\def\mdyn{$M_{\rm dyn}$}
\def\mphot{$M_{\rm phot}$}
\def\cnterm{[C/N]$_{\rm term}$}
\newcommand{\fig}[1]{Fig.\ \ref{#1}}
\newcommand{\Fig}[1]{Figure \ref{#1}}
\newcommand{\newtext}[1]{{\color{blue} #1}}
\def\gammaL{$\Gamma_{L}$}
\def\mhi{$M_{\rm hi}$}
\def\mlo{$M_{\rm lo}$}
\def\lhi{$L_{\rm hi}$}
\def\llo{$L_{\rm lo}$}
\def\lfin{$L_{\rm fin}$}

\def\llofit{$\log(L_{\rm lo}/L_\odot) = 4.39^{+0.10}_{-0.16}$}
\def\lhifit{$\log(L_{\rm hi}/L_\odot) = 5.20^{+0.17}_{-0.11}$}

\def\mlofit{$M_{\rm lo} = 6.9^{+0.9}_{-0.9}$\msun}
\def\mhifit{$M_{\rm hi} = 20.3^{+6.1}_{-2.6}$\msun}



\section{Introduction} \label{sec:intro}
Linking supernovae (SNe) to their progenitor stars is a powerful test of stellar evolutionary theory. The first SN for which this was possible, SN1987A, famously threw up the surprise of a blue progenitor \citep{Walborn87,Sonneborn87,Gilmozzi87}, whereas theory at the time predicted that such a star would explode as a Red Supergiant (RSG). Since then, it has been realised that SN1987A-like events are relatively rare, that SN1987A's progenitor likely had a complicated evolutionary history \citep[e.g.][]{Podsiadlowski92-87A}, and in fact the most common H-rich supernovae (classified as either II-P or II-L) do indeed have red progenitors \citep{Smartt09}. 

Beyond simply predicting the correct colour, one can also look at the luminosity distribution of II-P/L progenitors and test theoretical predictions by comparing to the expectations from population synthesis. Exclusively until now, such comparisons have involved converting the pre-explosion luminosities \lfin\ into initial masses \minit\ using stellar models, then comparing the inferred \minit\ distribution to a Salpeter initial mass function (IMF) with an upper and lower mass cutoff \citep{Smartt09, Smartt15, Davies-Beasor18}. In the first study of this kind, \citet[][ hereafter S09]{Smartt09} determined an upper mass cutoff for the progenitors of SNe II-P of \mmax=16\msun. Since evolutionary models at the time had stars with initial masses of up to 30\msun\ dying as RSGs, this tension -- and the fate of stars with \minit\ between 16-30\msun\ -- was termed the `Red Supergiant Problem'. The result has received a great deal of attention in the literature, due at least in part to the fact that this possible upper mass cutoff resonates with contemporaneous numerical work. Specifically, several independent authors have found that the likelihood of forming a black-hole at core collapse increases dramatically above masses of 16-20\msun\ \citep{OConnor-Ott11,Horiuchi14,Ertl16,Mueller16,Sukhbold18}, due to the transition from convective to radiative core carbon burning near the end of the star's life \citep{Sukhbold-Adams19}.

After the initial study by S09, a larger sample of progenitors was used by \citet[][ hereafter S15]{Smartt15} to revise the value of \mmax\ upwards to 17\msun. Further, \citet[][ hereafter DB18]{Davies-Beasor18} later revisited the complexities of converting a pre-explosion brightness to bolometric luminosity \lfin, and the conversion of \lfin\ to \minit, as well as other sources of systematic errors, and argued that \mmax\ was 21\msun\ once systematic errors had been taken into account, but with a 1-sigma upper error bar that extended up to 30\msun. The conclusion of DB18 was therefore that the evidence for a population of `missing' stars had only a minor statistical significance.

The studies of the RSG Problem mentioned above (S09, S15, DB18) tended to work in the mass-plane; that is, they infer initial masses for the progenitors, then compare these masses to the expectation of a stellar initial mass function (IMF) and a constant star-formation rate. The RSG Problem itself describes the tension between the inferred value of \mmax\ (16-21\msun) and its expectation value, commonly quoted as being 30\msun. Before continuing, it is worth emphasising that this expected value of \mmax\ is {\it not} a theoretical prediction; rather, it is the observed maximum luminosity of RSGs \lmax, also known as the Humphreys-Davidson Limit \citep[or H-D limit, ][]{Humphreys-Davidson79}, converted to an initial mass. The first measurements of the H-D limit placed it at $\log(L_{\rm max}/L_\odot) = 5.8$. This has subsequently been revised downwards to $\log(L_{\rm max}/L_\odot) = 5.5$ in the Magellanic Clouds \citep[][ hereafter DCB18]{DCB18}; and, as we will show in this paper, \lmax\ appears to be the same in the Milky Way. The value of \mmax\ is then simply the initial mass of the evolutionary track which terminates in the RSG phase with $L$=\lmax. According to the STARS evolutionary tracks used in S09, this corresponds to an initial mass of 27\msun, which is the basis for the expectation value of $\sim$30\msun\ quoted in reference to the RSG Problem. 

{Alternative techniques to study progenitor masses to SN IIP have been explored, but do not yet provide any additional reliable information. SN lightcurves have been modelled by several authors \citep[e.g.][]{Bersten11,Utrobin-Chugai17,Morozova18}, the shape and duration of which should yield information on the ejecta mass, which in turn should be related to the stellar initial mass. However, it has been shown that large degeneracies exist in the model parameter space which are difficult to break \citep[][]{Dessart-Hillier19,Goldberg19}. The progenitor's surface abundances, as studied from spectroscopy of the SNe at very early times, can also provide evidence as to the progenitor mass \citep[][]{Davies-Dessart19}, but this work has yet to be empirically road-tested. Finally, one may measure the mass of oxygen in the SN ejecta, which again should be a function of the progenitor mass \citep[][]{Jerkstrand14,Valenti16}. However, when exploited on SNe which also have pre-explosion imaging mass estimates, there is no apparent correlation between the two independent measurements, implying that either one or both are flawed (DB18). }

When attempting to infer the properties of the progenitor distribution, working in the mass-plane (i.e. beginning by converting \lfin\ to \minit\ via stellar models) serves to add in an extra layer of model dependence, and hence uncertainty, into the results. Furthermore, the use of a single set of evolutionary tracks neglects the complexity of `real' stellar evolution. For example, stars have a broad distribution of rotation rates \citep{Ramirez-Agudelo13};  they are often in multiples \citep{Sana12}, and so experience loss/gain of mass and/or angular momentum \citep[e.g.][]{Eldridge08}; and stellar models still rely on highly uncertain input physics such as mass-loss rates, convective mixing/overshooting \citep{Jones15}, plus any other form of `weather' that prevents two stars with similar bulk properties from following the exact same evolutionary path. All of these complicating factors are naturally accounted for by employing an {\it empirical} $L$-distribution with which to compare the observed luminosities of the SN progenitors. 


In S09, the sample of SN progenitors consisted of 20 events, with 7 detections (the rest being upper limits). Of these 7 detections, the most luminous was found to be SN1999ev with $\log(L/L_\odot) = 5.1\pm0.2$\footnote{The nature of the progenitor of SN1999ev, as well as others in the S09 sample, is now considered highly uncertain due to its apparent location in a star cluster \citep{Maund14}. The brightest progenitor in S09, as re-derived by DB18, is SN2006my with $\log(L_/L_\odot) = 4.97\pm0.18$.}. The absence of progenitors within the range $5.1 < \log(L_/L_\odot) < 5.5$, corresponding a mass range of 16-30\msun\ according to the STARS models in S09, was the basis for the original claim for `missing' progenitors. In the subsequent papers S15 and DB18, this sample size of detections grew to 14, the most luminous of which was SN2009hd with $\log(L/L_\odot) = 5.24\pm0.08$. In the following Section, we will argue that much of this tension be explained as being a consequence of a small sample. Later, in Section \ref{sec:Ldist} will also look at the $L$-distribution of Type-II progenitors to estimate the upper $L$ cutoff, which can be directly compared to the predictions from stellar evolution. Our findings are summarised in Sect.\ \ref{sec:conc}.

\begin{figure}
\begin{center}
\includegraphics[width=8.5cm]{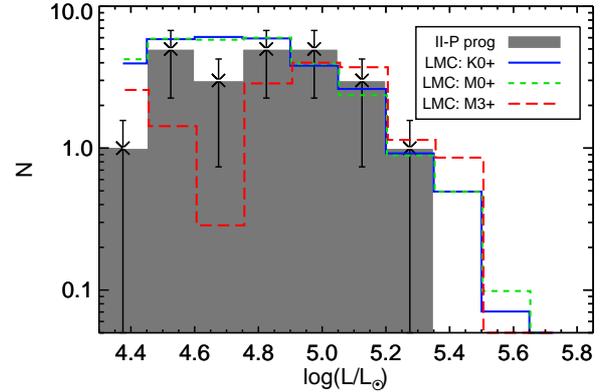}
\caption{The $L$-distribution of II-P progenitors is shown as the grey-filled histogram. Overplotted are the observed luminosity distributions of all RSGs (blue), all M supergiants (green) and all supergiants with spectral type M3 or later (red) in the Large Magellanic Cloud. Each LMC $L$-distribution has been rescaled to give the same number of objects with $\log(L_/L_\odot) > 4.8$ as the SN progenitors.}
\label{fig:Ldist}
\end{center}
\end{figure}

\section{A re-evaluation of the `missing' RSG progenitors} \label{sec:missing}

\subsection{The input sample}
Our sample of II-P and II-L progenitors comes from that of DB18, which itself was based on that in S15 but with the inclusion of SN2008cn. To this sample, we have added the following more recent events:

\begin{itemize}
\item SN2017eaw: the progenitor for this SN was studied in \citet{Kilpatrick-Foley18} and \citet{vanDyk19}, and was detected by Hubble Space Telescope {(\it HST)} in several bands. Both studies estimated foreground extinctions of $A_V = 1.0 \pm 0.1$, based on the strengths of the diffuse interstellar bands. Also, both attempted to obtain bolometric luminosities by modelling the SED, though both relied upon MARCS model atmospheres which are known to struggle to fit the optical and infrared spectra simultaneously \citep{rsgteff}, and did not have enough photometry in the mid-IR to constrain the emission from circumstellar dust. As an alternative estimate of \lbol, we take the F160W photometry, which is closest to the intrinsic peak of the SED and is less affected by extinction, and apply a bolometric correction typical of late-type RSGs (see DB18). Though BCs in the $H$-band were not discussed in DB18, analysis of the same dataset reveals that all RSGs in that study's sample have ${\it BC}_{\rm F160W} = 2.6\pm0.1$ irrespective of spectral type. The nature of the DB18 study is such that this BC {\it includes} the effect of circumstellar extinction and mid-IR excess. Using the pre-explosion brightness of $m_{\rm F160W} = 19.36\pm0.01$ and a distance to the host galaxy of NGC6946 of 7.72$\pm$0.32Mpc \citep[see discussion in][]{vanDyk19}, we find a terminal luminosity of $\log{L/L_\odot} = 4.96\pm0.11$. This is in very good agreement with the \citet{Kilpatrick-Foley18} and \citet{vanDyk19} studies.
\item SN2018aoq: the detection of this SN's progenitor was presented in \citet{ONeill19}. There were detections in the $V$, $I$ and $H$ bands with {\it HST}. The foreground reddening was determined from comparisons to of the lightcurve to other similar SNe, finding a value of \ebmv=0.03$\pm$0.01. Following our methodology for SN2017eaw, using a pre-explosion brightness of $m_{\rm F160W} = 21.89\pm0.02$ and a distance of  18.2$\pm$1.2kpc, we find a terminal luminosity of $\log{L/L_\odot} = 4.63\pm0.12$, which is somewhat lower than that of O'Neill et al. 
\end{itemize}

In addition, we have used the revised the distance to NGC6946, measured from the tip of the Red Giant branch \citep[7.72$\pm$0.32Mpc, see][]{vanDyk19}, for SN2002hh, SN2004et as well as SN2017eaw. We have not included the event SN2016cok ($\equiv$ASSASN-16fq), owing to the outstanding uncertainty over the progenitor \citep{Kochanek17}.

{The main criteria for inclusion in the sample is that it must be of type-II, be located in a galaxy with pre-explosion imaging, be near enough for the progenitor to be resolved (a distance limit of $\sim$30Mpc), and have either a progenitor detection or a meaningful upper limit. The sample likely has a bias towards higher metallicities (SMC-like or higher) since it is the more massive Local Group galaxies that have the more extensive archival imaging. Finally, there is potential bias in our sample of {\it detections} against objects with high line-of-sight reddening. Specifically, in the case of a non-detection, we may potentially underestimate the upper limit to the progenitor's luminosity if the object is obscured. In the case of obscuration by circumstellar extinction, in DB18 we attempted to correct for this bias by adopting empirical bolometric corrections based on observations of reddened late M-type supergiants. However, this does not account for extra extinction intrinsic to the host galaxy. \citet{Jencson17} have argued that there may be a hidden population of nearby core-collapse SNe which are hidden by large amounts of extinction intrinsic to the host galaxy, and that these objects may account for 10-20\% of the local supernova rate. It remains to be see if these `hidden' SNe have progenitors different to those discovered in the optical.}

\subsection{An empirical $L$-distribution of RSGs}
As our basis for comparison, we take the $L$-distribution of RSGs, defined as those with spectral types of K0 or later (K0+) in the Large Magellanic Cloud (LMC) as measured by DCB18. By studying the LMC rather than the Milky Way, we avoid issues such as uncertain distances and high interstellar reddening which can fatally affect the sample completeness. The LMC $L$-distribution in DCB18 was determined by searching, cross-matching and combining various multi-wavelength catalogues. By studying sources in the optical though mid-infrared, the authors were able to detect any sources with high circumstellar extinction that would previously have been missed in optical-only searches. Bolometric luminosities were determined by integrating under the observed SED, thus removing any dependence on uncertain bolometric corrections. The implicit assumption made was that any flux lost in the optical due to circumstellar extinction was re-radiated in the mid-IR. We found only one object where this assumption seemed to have been invalid (WOH G64), where a more detailed study of the SED by \citet{Ohnaka08} revealed a lower luminosity. Here, we have replaced the luminosity listed by DCB18 for this star by that determined by Ohnaka et al. {With this data-point corrected, the luminosity cutoff at $\log(L_{\rm max}/L_\odot) = 5.5$ is clear (see Fig.\ 2 in DCB18). Furthermore, population synthesis analysis in DCB18 has shown that any RSGs above this limit (as predicted by theory) are unlikely to have been missed. }

The majority of the SNe in the sample have metallicities between 12+$\log$(O/H) = 8.3-8.9, or LMC-like to slightly super-Solar (S09). Though we do not have a statistically complete $L$-distribution for the Galaxy, we will argue in Sect.\ \ref{sec:MW} that there is no evidence for the Humphreys-Davidson limit \lmax\ being brighter in the Galaxy than in the LMC. The benefit of using an empirical $L$-distribution, as opposed to a theoretical one, is that we automatically bypass the uncertainties in stellar evolution, some of which (contribution of post-binary interaction objects, initial rotation rate distribution, magnetic fields) are almost intractable. For example, the relation between initial mass and terminal luminosity has been shown to have a great deal of dispersion once binarity is taken into account  (Zapartas et al., submitted).
 
In \fig{fig:Ldist}, we plot the luminosity function of all cool supergiants in the LMC from DCB18. It is important to note that our empirical sample of RSGs contains stars at {\it all stages} of RSG evolution, whereas the SN progenitors are exclusively RSGs at the very end of the phase. In DB18 we argued that these late-evolution RSGs tend to have later spectral types (M3 or later), based on two forms of evidence. Firstly, in star clusters with large numbers of RSGs, we see the more evolved stars having later spectral types. Secondly, SNe with multi-colour pre-explosion photometry invariably show that the progenitor was very red, consistent with having a spectral type of M3 or later (e.g. SN2003gd, SN2004et). Therefore, in an attempt to isolate the LMC RSGs which are closest to SN, we have made cuts on spectral type of M0 or later (which we call M0+) and M3 or later (M3+). 

In \fig{fig:Ldist} we overplot the $L$-distributions of the master sample of RSGs in the LMC, as well as the two subsamples. Each histogram has been normalised to reproduce the total number of objects as the II-P progenitors within the range $4.8 < \log(L/L_\odot) < 5.6$, below which the LMC RSG sample is likely incomplete (DCB18). No matter how we slice the sample, in all cases we see a definite cutoff at $\log(L_{\rm max}/L_\odot) \simeq 5.5$. In the total (K0+) sample and the M0+ subsample, we see overall luminosity distributions which are very similar in shape, and which match the observed II-P distribution rather well. In the late-type subsample (M3+) we see the distribution roll-over to smaller numbers below a brightness of $\log(L/L_\odot) \simeq 4.9$. This is likely an effect of incompleteness -- fainter, redder objects are more likely to be missing from the DCB18 sample -- though it could also be explained as being caused by fainter (i.e. lower initial mass) RSGs spending less time at later spectral types as a fraction of their overall RSG lifetime. The consequences of these two possible explanations for our conclusions are discussed in the next section. 

\begin{figure}
\begin{center}
\includegraphics[width=8.5cm]{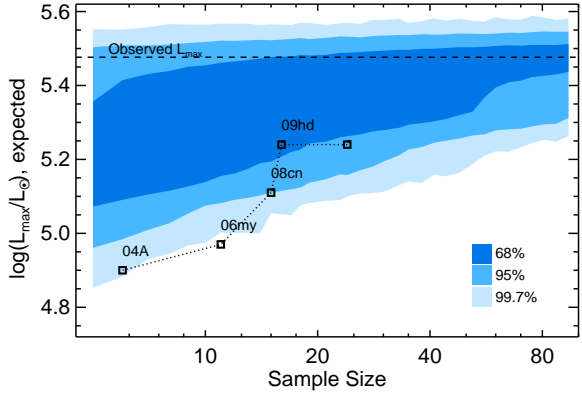}
\caption{The expected luminosity of the brightest supernova progenitor for a range of sample sizes. The larger the sample, the higher the probability that the brightest progenitor has a luminosity close to the intrinsic \lmax\ (shown by the blue dashed line). The shaded blue regions indicate the confidence limits on \lmax\ as indicated in the legend. The data points show the observed evolution of $L_{\rm max, prog}$ as the sample size has increased over time.}
\label{fig:Lss}
\end{center}
\end{figure}


\subsection{The brightest expected supernova progenitor}
Under the assumption that the $L$-distribution of late-M supergiants shown in the previous section (labelled M3+) effectively describes that of II-P/L supernova progenitors, the probability of a given SN progenitor having a luminosity $L$ can be found simply by randomly sampling this $L$-distribution. Obviously, the IMF combined with the shorter lifetime of more massive stars dictate that one is more likely to find a faint progenitor than a bright one, and the probability of finding a bright progenitor increases with increasing sample size. That is, the larger the sample size, the brighter we expect the luminosity of our brightest SN progenitor $L_{\rm max, prog}$ to be. The $L$-distribution in \fig{fig:Ldist} allows us to quantify this. 

We determine the probability distribution function (PDF) of $L_{\rm max, prog}$ for a sample size $N$ by performing a simple Monte-Carlo (MC) experiment. We first randomly sample the luminosities of $N$ M-supergiants in the LMC with spectral types M3 or later, where each star's luminosity is itself sampled randomly from a Gaussian distribution with a standard deviation equal to the star's 1$\sigma$ error. In each trial, we determine the brightest of the $N$ progenitors, which we set as $L_{\rm max, prog}$ for that trial. We then repeat $10^4$ times for each value of $N$ to determine the PDF of $L_{\rm max, prog}$ at that $N$. We perform this same experiment for a range of sample sizes. 

The results of this experiment are shown as the shaded region in \fig{fig:Lss}, where the different colours indicate the confidence limits in the legend. As expected, the plot demonstrates that the smaller the sample size, the fainter the $L_{\rm max, prog}$ one expects to observe. Specifically, at sample sizes below 10 one expects to find $\log(L_{\rm max, prog}/L_\odot) \simeq 5.2$, but with a large dispersion. At sample sizes greater than 80, the absence of a progenitor brighter than $\log(L_{\rm max, prog}/L_\odot) \simeq 5.25$ starts to become significant at the 3$\sigma$ level.

In order to compare how the observed $L_{\rm max, prog}$ has evolved as sample size has grown, in \fig{fig:Lss} we overplot the brightest SN progenitor as a function of the sample size at the time that SN was observed. For small sample sizes (<10), the disagreement with the expectation appears to be significant. However, as the sample size grows, the observed trend (black squares) starts to follow the results of our MC experiment more closely. Presently, the discrepancy between the luminosity of the brightest SN progenitor to date (SN2009hd) and the observed H-D limit has a significance of 1-2$\sigma$. However, this does not take into account the error on SN2009hd, which had a luminosity of $\log(L/L_\odot) = 5.24 \pm 0.08$, or indeed that of SN2009kr ($\log(L/L_\odot) = 5.13 \pm 0.23$) which was less than 2$\sigma$ from the H-D limit at $\log(L/L_\odot) = 5.5$. 

In the previous section we noted that the luminosity distribution of the M3+ RSGs is different to that of the total RSG sample. We also offered two explanations for this -- one physical, that it is caused by relatively shorter durations of the M3+ phase for lower mass RSGs; and one systematic, that it is caused by greater statistical incompleteness at lower luminosities for the M3+ subsample. If the latter explanation is correct, it would introduce a bias into our results. Specifically, by randomly sampling from a population that had too few faint objects, we would artificially increase the likelihood of selecting a bright progenitor in each MC trial. Correcting for this bias, should it exist, would effectively pull the blue shaded region in \fig{fig:Lss} down to lower luminosities, decreasing further the statistical significance of the RSG problem.

\begin{table*}
\setlength{\extrarowheight}{6pt}
\caption{Observed properties of luminous Milky Way RSGs.}
\begin{center}
\begin{tabular}{lcccccc}
\hline \hline
Star & RA & DEC & $D$/kpc & $D$ method & E(B-V) & \logl \\
\hline
S Per & 02 22 51.7 & +58 35 11.5 & 2.42$^{+0.11}_{-0.09}$ & Maser & 0.69$^{+0.16}_{-0.51}$ & 5.09$^{+0.08}_{-0.15}$ \\
VY CMa & 07 22 58.3 & -25 46 03.2 & 1.20$^{+0.13}_{-0.10}$ & Maser & 0.06$^{+0.05}_{-0.05}$ & 5.25$^{+0.09}_{-0.08}$ \\
CK Car & 10 24 25.4 & -60 11 29.0 & 2.92$^{+0.19}_{-0.15}$ & OB & 0.30$^{+0.15}_{-0.18}$ & 5.20$^{+0.10}_{-0.10}$ \\
RT Car & 10 44 47.1 & -59 24 48.1 & 2.38$^{+0.18}_{-0.15}$ &  Maser & 0.53$^{+0.12}_{-0.12}$ & 5.11$^{+0.10}_{-0.09}$ \\
EV Car & 10 20 21.6 & -60 27 15.8 & 2.96$^{+0.22}_{-0.20}$ & OB & 0.40$^{+0.12}_{-0.29}$ & 5.46$^{+0.10}_{-0.14}$ \\
VX Sgr & 18 08 04.0 & -22 13 26.6 & 1.56$^{+0.11}_{-0.10}$ & Maser & 0.26$^{+0.40}_{-0.19}$ & 5.44$^{+0.15}_{-0.09}$ \\
BC Cyg & 20 21 38.5 & +37 31 58.9 & 1.71$^{+0.04}_{-0.04}$ & OB & 0.80$^{+0.80}_{-0.60}$ & 5.31$^{+0.25}_{-0.14}$ \\
RW Cyg & 20 28 50.6 & +39 58 54.4 & 1.62$^{+0.04}_{-0.04}$ & OB & 0.70$^{+0.40}_{-0.50}$ & 5.10$^{+0.16}_{-0.17}$ \\
NML Cyg & 20 46 25.5 & +40 06 59.4 & 1.61$^{+0.13}_{-0.11}$ & Maser & 0.50$^{+0.25}_{-0.25}$ & 5.36$^{+0.07}_{-0.07}$ \\
$\mu$ Cep  & 21 43 30.5 & +58 46 48.2 & 0.94$^{+0.14}_{-0.04}$ & OB & 0.46$^{+0.09}_{-0.09}$ & 5.43$^{+0.15}_{-0.07}$ \\
MY~Cep & 22 54 31.7 & +60 49 39.0 & 3.00$^{+0.35}_{-0.29}$ & OB & 2.04$\pm$0.07 & 5.11$^{+0.08}_{-0.07}$\\
PZ Cas & 23 44 03.3 & +61 47 22.2 & 2.81$^{+0.22}_{-0.19}$ & Maser & 0.51$^{+0.40}_{-0.11}$ & 5.36$^{+0.19}_{-0.09}$ \\
\hline
\end{tabular}
\end{center}
\label{tab:obs}
\end{table*}%

\subsection{Comparisons to previous work}
Our result that the statistical significance of the `RSG Problem' is below 2$\sigma$ seems to contradict the conclusions of S09 and S15. These authors estimated the significance of the `missing' RSGs to be in the region of 4$\sigma$, based on their sample size and the lack of objects with luminosities between \lmax\ and that of their brightest progenitor \lhi. However, we note that S09/S15 did not seem to account for the uncertainties on the luminosities of the progenitors when determining this (see previous section). 

We can see from \fig{fig:Lss} that when the sample size was only $\sim$10, as it was in S09, there was the appearance that the result was significant. However, this was before the events with brighter progenitors (SN2008cn, SN2009kr and SN2009hd). Though SN2009kr and SN2009hd were included in S15, both had their luminosities revised upwards by DB18, who tuned up the assumptions about the bolometric corrections and foreground extinctions. Indeed, in DB18 we argued that the significance of the `missing' stars was lower than claimed in S09 and S15, due to the additional uncertainties in the theoretical \minit-\lfin\ relation, plus the small sample size.   Therefore, the conclusions of this part of our current study are consistent with those of DB18.



\begin{figure}
\begin{center}
\includegraphics[width=8.5cm]{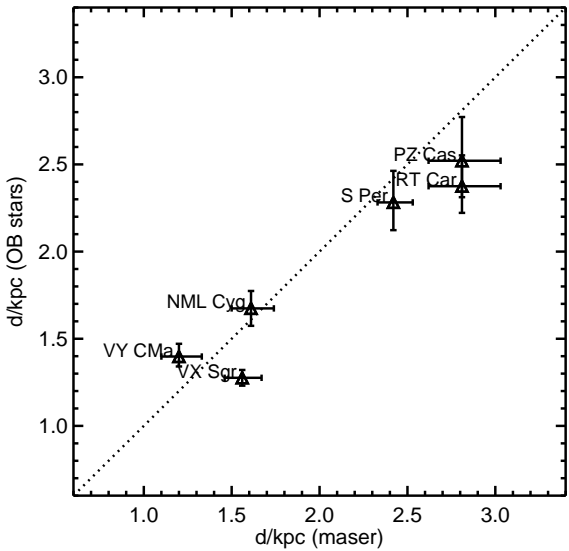}
\caption{Comparison between the distances obtained from maser parallaxes (x-axis) and those from the Gaia DR2 parallaxes of neighbouring OB stars (y-axis). The dotted line shows the 1:1 correlation. }
\label{fig:distcomp}
\end{center}
\end{figure}

\subsection{Is the $L$-distribution of RSGs different in the Milky Way?} \label{sec:MW}
The most obvious criticism one might make of this current work is that we employ an empirical $L$-distribution measured in the LMC, whereas many of the SNe in the current sample occurred in galaxies where the metallicity is thought to be closer to Solar (S09). A statistically complete sample of RSGs in the Milky Way does not exist at the time of writing. The construction of such a sample would be extremely difficult, as it would have to overcome the obstacles of high foreground extinction beyond distances of $\sim$3kpc, contamination by foreground red giants and AGB stars, as well as uncertain distances. However, we {\it can} investigate one key feature of the Galactic RSG $L$-distribution; specifically the brightness of the H-D limit \lmax. 

In S09, the expectation value of \lmax\ was determined by comparing to the brightest known Galactic RSGs in \citep{Levesque05}. These bright RSGs were some of the more famous members of the class, such as VY~CMa and $\mu$~Cep. The luminosities of these stars was determined by Levesque et al.\ by two methods; the extrapolation of model fits to the optical spectral energy distribution (SED), and from bolometrically-corrected $K$-band brightnesses. Distances were inferred by assuming that the RSGs were physically associated with the nearest OB association, and adopting the appropriate distance from the literature. Here, we now re-appraise the luminosities of the brightest of these RSGs. To do this, we firstly re-evaluate their distances and foreground extinctions, then determine their bolometric luminosities by integrating under their observed SEDs from the optical to the mid-infrared. 

\subsubsection{Distances and reddenings}
To determine distances to the Galactic RSGs, we employ parallax measurements. Optical parallaxes of RSGs, such as those measured by Gaia \citep{GaiaDR2}, are extremely problematic due to the stars' inhomogeneous surfaces and the fact that their sizes are often comparable to the baseline for the parallax measurement (i.e.\ the size of the Earth's orbit around the Sun) \citep[see e.g.][]{Chiavassa11gaia}. Fortunately, there are other ways to determine RSG distances from parallaxes. Firstly, one may employ radio parallax measurements of circumstellar masers. Secondly, one may take the average parallax of the neighbouring OB stars, under the assumption that the RSG is a part of the same association \citep[see also][]{Humphreys78,Levesque05}. Where possible, we employ both methods here to verify the accuracy of the results. 

To measure reddenings we can again look at the neighbouring OB stars, and use the colour excess to estimate the average line-of-sight extinction to the RSGs. Though there may well be extra extinction to each RSG due to circumstellar material, we can still obtain a bolometric luminosity by integrating the full SED under the assumption that any flux lost at short wavelengths is re-radiated in the infrared (see also DCB18). 

For each RSG in our sample, we search the SIMBAD database for OB stars with known spectral types within 30\arcmin\ of the star. Where there are a large number of OB stars, we take the nearest 50. From spectral types of the OB stars, we obtain the intrinsic $B-V$ colours from \citet{martins-plez06} and \citet{Fitzgerald70}, and use the observed colours to determine $E(B-V)$ for each OB star. We then take the sigma-clipped mean of these reddenings, clipping at 2$\sigma$. 

The parallax measurements for the OB stars were obtained from Gaia DR2 \citet{GaiaDR2}. We again performed sigma-clipping with a threshold of 2$\sigma$, then took the sigma-weighted mean of the remaining stars. This mean parallax was converted to a distance following the procedure described in \citet{DB-distances}. The comparisons of these `OB star' distances to those obtained from maser parallaxes are illustrated in \fig{fig:distcomp}. The data points all follow the 1:1 line, implying that both methods are consistent with one another. Formally, analysis of the residuals reveals a mean offset of 150$\pm$240\,pc. The standard deviation on the offset likely represents the absolute precision on the OB star method, since the OB associations and complexes that host the RSGs could easily be of order $\sim$100pc.

\subsubsection{Luminosities}
Bolometric luminosities are determined by first collating broad-band photometry for each RSG spanning the optical to mid-infrared. The brightnesses of these stars makes this a non-trivial task, since many are often saturated in contemporary surveys. We took optical photometry from Gaia DR2 \citep{GaiaDR2}, SDSS-IV DR15 $i$-band \citep{SDSS-15} and \citet{Morel78}, near-IR photometry from 2MASS \citep{2MASS} and \citet{Morel78}, and mid-IR photometry from IRAS and MSX \citep{IRAS,MSX}. Having several survey sources overlapping in wavelength permits the identification and rejection of spurious photometric points, due to e.g. saturation. For each star, the good photometric data are dereddened and interpolated in the $\log$(Flux) and $\log(\lambda)$ plane, then integrated to find the bolometric luminosity. Flux shortward of $B$ is assumed to contribute a negligible amount to the bolometric flux (see also DCB18).

\subsubsection{Results}
Our revised distances, reddenings and bolometric luminosities of the bright Galactic RSGs are listed in table \ref{tab:obs}. From the last column, it can be seen that no star has a luminosity greater than $\log(L/L_\odot) = 5.5$. In terms of how these values compare to previous estimates, all are roughly consistent with the bolometrically corrected $K$-band estimates of \citet{Levesque05}. Any differences with the values listed in \citet{Mauron-Josselin11}, who like us obtained \lbol\ by integrating under the SED, can be attributed to changes in distance and foreground reddening. The most notable change is for the star EV~Car, which in \citet{Mauron-Josselin11} had a distance of 4.2kpc. We investigated all known OB stars with spectral types earlier than B3 in the whole Carina star forming region, spanning 10\degr\ in Galactic longitude, and consistently found distances between 2.4-3.0kpc. 

Though we list only 12 stars in Table \ref{tab:obs}, we emphasise that these have been previously identified as the brightest optically identified RSGs in \citet{Levesque05} and  \citet{Mauron-Josselin11}. The objects in Table \ref{tab:obs} are therefore only a subset of optically-identified Milky Way RSGs. The full sample studied by \citet{Levesque05} contained over 80 stars, which was by no means complete. A search of SIMBAD reveals that there are over 100 stars in the Galaxy with optically-identified spectral types later than K0 and luminosity classes Ib or brighter. With a sample size this large, one can expect the brightest observed RSG to be sampling close to \lmax\ (cf.\ \fig{fig:Lss}). We therefore consider the lack of any known RSGs in the Milky Way with luminosities above $\log(L/L_\odot) = 5.5$ evidence that the H-D limit at Solar metallicity is similar to that found in the Magellanic Clouds.

\section{A re-analysis of the $L$-distribution of II-P/L progenitors, and the upper cutoff in luminosity and mass} \label{sec:Ldist}
In the previous section we discussed the statistical significance of the `missing' RSG progenitors by comparing the luminosities of the brightest II-P progenitors with that of the RSG luminosity limit \lmax. However, we may learn more about the properties of II-P progenitors by studying the observed progenitor sample as a whole. In this Section, we analyse the observed luminosity distribution of II-P/L progenitors to evaluate the properties of the parent population. In particular, the aim is to provide the most robust measurement to date of the upper luminosity boundary \lhi. This in turn can be used to infer the upper mass boundary \mhi, allowing us to clearly separate the random errors caused by the finite sample size from the systematic errors introduced by the theoretical mass-luminosity relation. 

To date, analysis of the underlying population of II-P/L progenitors has focused on the inferred masses, fitting an analytical function based on the IMF and upper/lower mass limits (S09, S15, DB18). There are two problems with this analysis, discussed earlier in this work as well as in DB18. Firstly, it implicitly assumes that the highest mass progenitor in the sample will have a mass close to \mhi, which in a finite sample causes one to underestimate \mhi. Secondly, converting luminosities to masses injects all model errors and uncertainties into the analysis (see discussion in Sect.\ \ref{sec:intro}). 

Here, to circumvent these problems, we present a new analysis method. Firstly, we study the luminosity distribution, rather than the mass distribution. Secondly, we employ a Monte-Carlo (MC) method to randomly sample from a master population, which simulates the effects of a finite and small sample, which we now describe in more detail.


\begin{figure}
\begin{center}
\includegraphics[width=8.5cm]{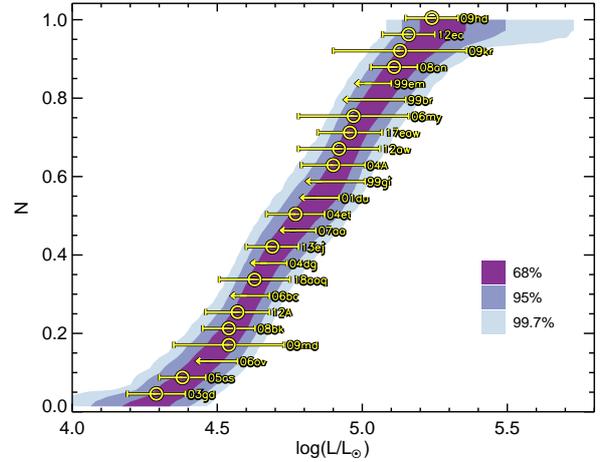}
\caption{The cumulative luminosity distribution (CLD) of SN progenitors. The individual observations, sorted in order of increasing $L$, are shown in yellow. The shaded contours show the confidence limits of the underlying CLD, as determined by the Monte-Carlo experiment described in the text. }
\label{fig:lcumobs}
\end{center}
\end{figure}



\subsection{The observed cumulative luminosity distribution of SN progenitors}
The first step is to create a proper description of the progenitor cumulative luminosity distribution (CLD) for our $N$ SN progenitors. Simply plotting the observed progenitors in order of increasing luminosity (such as in Fig.\ 6 in S15 or Fig.\ 5 in DB18) does not take into account that the errors on $L$ also affect the ranking. That is, perturbing the $L$ of a progenitor to a higher value would also cause that progenitor to be higher in the ranking, and vice-versa. 

We obtain a probability density distribution of the CLD by performing a simple Monte Carlo procedure. In each MC trial, we randomly sample from the probability distribution of each progenitor's pre-explosion brightness, distance, extinction, and bolometric correction to determine that progenitor's luminosity, then order the progenitors in increasing $L$. The probability distribution of the $n$th progenitor's luminosity $P_n (L)$ is determined from the luminosities of 30,000 MC trials. Note that $n$ does {\it not} correspond to an individual progenitor, but is the most likely $n$th brightest out of $N$ progenitors given the observational errors on all objects in the sample. 

The progenitor CLD is shown as the filled contours in \fig{fig:lcumobs}. Also shown on the this plot are the progenitor measurements in order of increasing $L$, to demonstrate the subtle differences between the two. Specifically, the CLD is narrower in the mid-range than the individual errors on the data-points, but extends to lower/higher $L$ at the bottom/top respectively. 

\begin{figure*}
\begin{center}
\includegraphics[width=17cm]{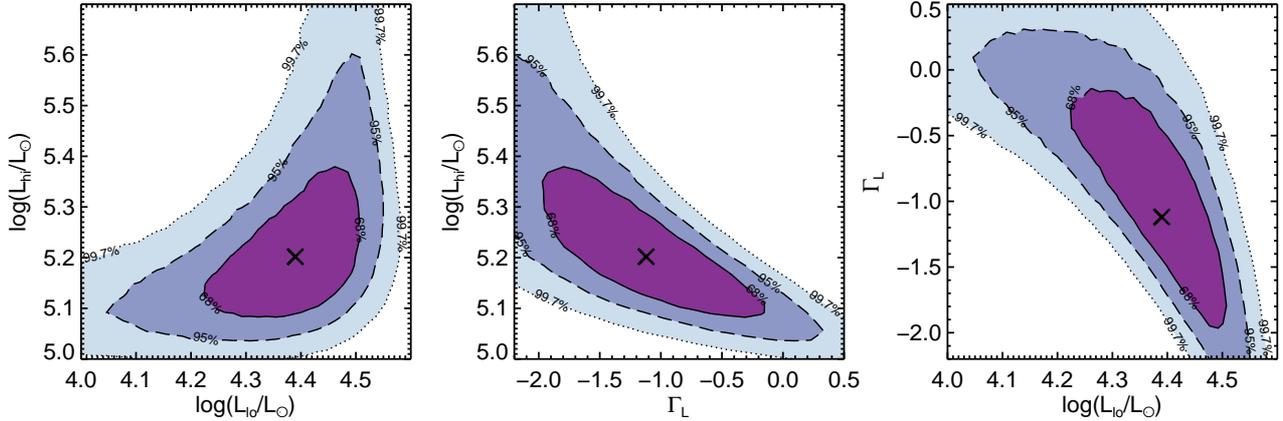}
\caption{2-D probability distributions from the CLD fitting, plotted as pairs of parameters for the three parameters in our grid. The best-fitting parameters are indicated by the cross in each panel. }
\label{fig:3panel}
\end{center}
\end{figure*}

\subsection{A grid of model cumulative luminosity distributions}
Next, we construct a model grid of CLDs to compare to the observed CLD. The CLD is modelled as a simple power-law, and therefore has three input parameters; the lower and upper luminosity cutoffs \llo\ and \lhi, and the power-law exponent \gammaL. At a given set of \{\llo, \lhi, \gammaL\} we again determine the CLD using Monte Carlo. We randomly sample $N$ luminosities from the distribution, and add on random noise according to the error bars of the observed $L$ distribution. For example, for the faintest progenitor in our simulated sample, we apply the same random error as for the faintest observed progenitor. We again repeat over 30,000 MC trials to determine the posterior probability distribution for the $n$th brightest progenitor out of $N$ objects.

\subsection{Finding the best fitting model}
At each point in the grid, we evaluate the best fitting model via a maximum likelihood analysis. The probability $P(n)$ that a model with parameters \{ \llo, \lhi, \gammaL \} reproduces the luminosity $L$ of the $n$th progenitor is, 

\begin{equation}
\displaystyle
P(n) = \sum_{L} \left( P_{\rm obs, \it n}(L) \times P_{\rm mod, \it n}(L) \right)
\end{equation}

\noindent where $P_{\rm obs, \it n}$ is the observed probability density as a function of luminosity for the $n$th progenitor (i.e. a horizontal row in \fig{fig:lcumobs}), and $P_{\rm mod, \it n}$ is the same but for the model.  The likelihood $\mathcal{L}$ that model \{\llo,\lhi,\gammaL\} fits the observed CLD is then,

\begin{equation}
\ln \mathcal{L} = \sum_n \ln P(n) 
\end{equation}

\noindent To compare the quality of fits of neighbouring models, the likelihoods are converted to $\chi^2$ values via, $\chi^2 = -2 \ln \mathcal{L}$. The best fitting model is defined to be that with the lowest $\chi^2$ value, while the models within the 68\%, 95\% and 99.7\% confidence limits are those $\chi^2$ values within 3.53, 8.02 and 14.16 of the minimum\footnote{These confidence intervals are defined for the 3 degrees of freedom of our model, following \citet{Avni76}.}, respectively.

\subsection{Results \& Discussion}
\begin{figure}
\begin{center}
\includegraphics[width=8.5cm]{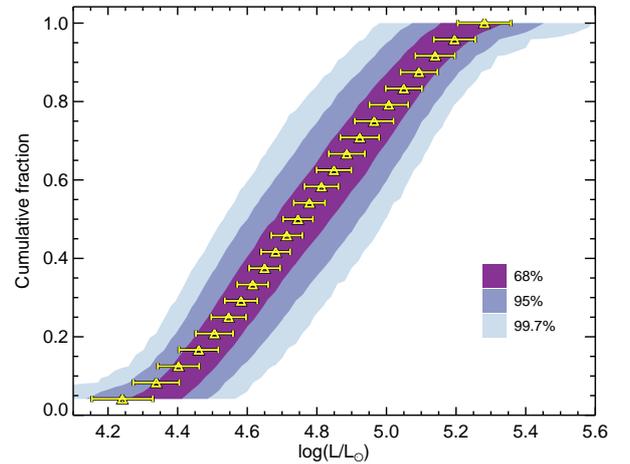}
\caption{The observed CLD (yellow points and error bars, corresponding to the 68\% limits in \fig{fig:lcumobs}) and the best fitting model CLD (shaded contours). }
\label{fig:cldbest}
\end{center}
\end{figure}

The probability distributions of each model parameter are illustrated in \fig{fig:3panel}, while the best-fitting model CLD is shown in \fig{fig:cldbest}. The best fit model parameters, when all are allowed to be free, are:

\begin{tabular}{rcc}
$\log(L_{\rm lo}/L_\odot)$ & = & $ 4.39^{+0.10}_{-0.16}$ \vspace{2mm}\\
$\log(L_{\rm hi}/L_\odot)$ & = & $ 5.20^{+0.17}_{-0.11}$ \vspace{2mm}\\
          $\Gamma_{\rm L}$ & = & $-1.12^{+0.95}_{-0.81}$ \vspace{2mm}\\
\end{tabular}

\noindent where the quoted errors are the 68\% confidence limits. The optimized value of \lhi\ is somewhat below the H-D limit (consistent with the `Red Supergiant Problem'), however the upper error bar stretches to quite high luminosities. Specifically, the tension with the observed H-D limit at $\log(L/L_\odot) = 5.5$ is within the 95\% confidence limit, analogous to a significance of less than 2$\sigma$.  The large upper error bar on \lhi\ is caused in part by the the degeneracy with \gammaL -- for steeper power laws, bright progenitors are expected to be rarer, meaning there is a stronger bias towards the highest observed luminosity progenitor being well below the intrinsic \lmax. The likelihood of the brightest progenitor in a finite sample having a luminosity close to \lmax\ is worse for steeper \gammaL\ (see centre panel of \fig{fig:3panel}).  A similar degeneracy exists between \llo\ and \gammaL. A steeper power-law slope forces the CLD to be closer to vertical at the faint end, which pulls \llo\ to higher values (see right-hand panel of \fig{fig:3panel}). 

\begin{figure}
\begin{center}
\includegraphics[width=8.5cm]{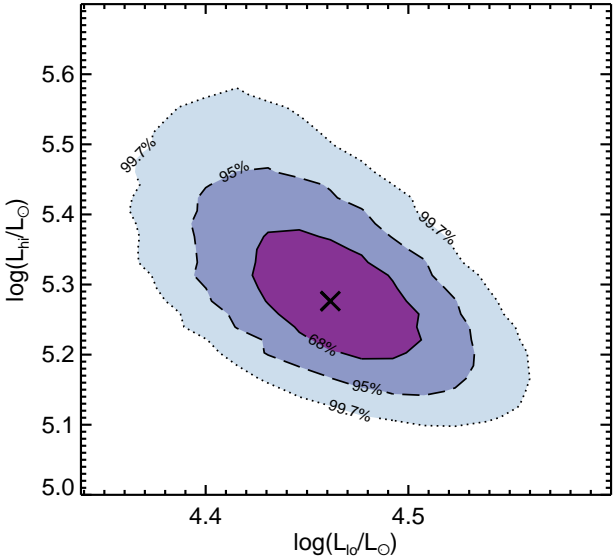}
\caption{Probability density as a function of \llo\ and \lhi\ when the power-law slope is constrained to \gammaL=-1.675, which is the expected value for this parameter when assuming a Salpeter IMF and a MLR of \lfin$\propto$\minit$^2$.}
\label{fig:L-2D-G}
\end{center}
\end{figure}

Given these degeneracies with \gammaL, it makes sense to look further into what the expectation value of the power-law slope might be. For an initial mass function $dN/dM \propto M^\Gamma$, and an initial mass -- final luminosity relation (MLR) that scales as $L \propto M_{\it init}^\alpha$, it can be shown that we would expect a luminosity function $dN/dL \propto L^{\Gamma_{\rm L}}$ with \gammaL$=(1-\alpha+\Gamma)/\alpha$. For a Salpeter IMF slope of $\Gamma=-2.35$, and a MLR slope of $\alpha=2$\footnote{All single-star evolutionary models we investigated (Geneva/STARS/Kepler) had $\alpha$ between 1.9 and 2.1.}, we would expect \gammaL=-1.675. Our best-fit value of \gammaL\  is shallower than this, but is within the $1\sigma$ confidence limits. If we constrain \gammaL=-1.675, we find the 2-D probability density function between \llo\ and \lhi\ shown in \fig{fig:L-2D-G}. Here, the optimal value of \lhi\ shifts to a higher value of $\log(L_{\rm hi}/L_\odot) = 5.28^{+0.12}_{-0.06}$, though the tension with the H-D limit remains $\sim 2 \sigma$.

\begin{table}
\setlength{\extrarowheight}{6pt}
\caption{Upper and lower limits to the progenitor mass ranges, with 95\% confidence intervals, for three different stellar evolution models. }
\begin{center}
\begin{tabular}{lcc}
\hline\hline
Model & \mlo/\msun & \mhi/\msun \\
\hline
   Geneva-nr &  7.8$^{+1.0}_{-1.3}$ & 20.5$^{+4.5}_{-2.5}$ \\
    Geneva-r &  6.4$^{+0.9}_{-1.2}$ & 17.9$^{+4.2}_{-2.3}$ \\
      KEPLER &  7.7$^{+1.0}_{-1.3}$ & 20.3$^{+4.5}_{-2.5}$ \\
       STARS &  7.1$^{+0.9}_{-1.2}$ & 17.9$^{+3.7}_{-2.1}$ \\
        MIST &  7.0$^{+1.0}_{-1.3}$ & 19.5$^{+4.6}_{-2.5}$ \\
\hline
\end{tabular}
\end{center}
\label{tab:masses}
\end{table}

\subsection{Conversion of \lhi\ to \minit}
With posterior probabilities for \llo\ and \lhi, we can convert these grids from luminosity-space to mass-space by simply using the MLRs of various stellar models. In Table \ref{tab:masses} we list the best fit mass ranges and 68\% confidence intervals for several single-star evolution models: Geneva \citep[both rotating and non-rotating, ][]{Ekstrom12}, KEPLER \citep{Woosley-Heger07}, STARS \citep{Eldridge08}, and MIST \citep{MIST}. Of these models, we note that the MIST and Geneva models are only evolved as far as the end of C-burning, and so may not accurately reproduce the properties within a few years of SN. Furthermore, we caution that we are only considering single star evolution, and that the MLR may have a different slope and a large degree of variance once binary evolution is included (Zapartas et al., in prep).

Using the STARS models, we find $M_{\rm lo} = 7.1^{+0.9}_{-1.2}$\msun\ and $M_{\rm hi} = 17.9^{+3.7}_{-2.1}$\msun, again quoting 68\% confidence limits, consistent with the 19\msun\ (+2\msun\ systematic) that we found in DB18. Mass limits for other suites of stellar models are listed in Table \ref{tab:masses}. In general the results show agreement for \mhi\ within the range 17.9-20.5\msun, implying that perhaps the best-fit is \mhi=19.2\msun\ with a systematic error of $\pm1.3$\msun. However, the random error is much larger, with typical 68\% confidence limits of +4.5 / -2.5\msun. Though again these numbers do not contradict the existence of the `RSG Problem', we caution that at present the significance of this result is within $2\sigma$. 

\section{Conclusions} \label{sec:conc}
We have taken a fresh look at the Red Supergiant Problem in two ways. Firstly, we compared the observed luminosities $L$ of II-P/L SN progenitors to those of late-type RSGs in the field, to assess whether we could reasonably expect to have sampled close to the upper luminosity boundary (i.e the Humphreys-Davisdon limit, or H-D limit, at $\log(L/L_\odot$=5.5) given the current sample size of progenitors. This is in contrast to previous studies, which convert the luminosities to masses, then compare to the expected mass distribution given the stellar initial mass function (IMF) and a theoretical estimate of the initial mass -- final luminosity relation (MLR). By looking at the problem in this new way we are able to eliminate the uncertainties and model-dependencies that are introduced when one first converts the pre-explosion luminosities to initial mass before comparing to expectation values. Our results indicate that the difference in $L$ between the brightest progenitor and the H-D limit of evolved RSGs can be explained in part by the small sample size. Quantitatively, the statistical significance of the RSG Problem is below 2$\sigma$.

Secondly, we model the observed $L$-distribution of II-P/L progenitors as a simple power-law with a slope \gammaL, and upper/lower luminosity limits \lhi\ and \llo. Our best-fit value of the upper luminosity boundary is \lhifit, again lower than the observed H-D limit, but again with a statistical significance of less than 2$\sigma$. The lower luminosity cutoff is measured to be \llofit. Our best fit value for the slope of the power law (\gammaL) is somewhat shallower than would be expected for a Salpeter IMF and a MLR of $L_{\rm fin} \propto M_{\rm init}^2$, though again the significance of this is low ($<1 \sigma$).  When forcing \gammaL\ to be the expected value of -1.675, we find the upper luminosity cutoff increases to $\log(L_{\rm hi}/L_\odot) = 5.28^{+0.12}_{-0.08}$, but the statistical significance of the tension with the H-D limit is still within $\sim 3\sigma$.

The limits on the terminal luminosities of II-P and II-L progenitors can be translated to initial masses using evolutionary models. By comparing several single-star codes, we find a lower mass limit between of \mlo=6-8\msun\ and an upper mass limit \mhi=18-20\msun. The 68\% confidence limits on \mhi\ are (+4.5, -2.3)\msun. The best fit value on \mhi\ is therefore very close to the prediction of various evolutionary codes that stars with initial masses greater than 20\msun\ form black-holes at core-collapse. However, we caution that the empirical uncertainties on \mhi\ are still very large, a situation that will not change until the sample size is at least doubled. 

\section*{Acknowledgements}
We thank Nathan Smith for helpful discussions, and the organisers of the FOE2019 meeting which inspired this work. We made use of the IDL astronomy library, available at {\tt https://idlastro.gsfc.nasa.gov}. EB is supported by NASA through Hubble Fellowship grant HST-HF2-51428 awarded by the Space Telescope Science Institute, which is operated by the Association of Universities for Research in Astronomy, Inc., for NASA, under contract NAS5-26555.




\bibliographystyle{mnras}
\bibliography{/Users/astbdavi/Google_Drive/drafts/biblio} 



\bsp	
\label{lastpage}
\end{document}